\documentclass[aps,preprint]{revtex4}
\usepackage{amsfonts}
\usepackage{amsmath}
\usepackage{amssymb}
\usepackage{subcaption}
\usepackage{graphicx}
\usepackage{grffile}
\usepackage{epstopdf}
\usepackage{float}

\begin{document}
\title{Joule-Thomson Expansion of RN-AdS Black Hole Immersed in Perfect Fluid Dark Matter }
\author{Yihe Cao}
\email{yihe@stu.scu.edu.cn}
\author{Hanwen Feng}
\email{fenghanwen@stu.scu.edu.cn}
\author{Wei Hong}
\email{thphysics_weihong@stu.scu.edu.cn}
\author{Jun Tao}
\email{taojun@scu.edu.cn}
\affiliation{Center for Theoretical Physics, College of Physics, Sichuan University, Chengdu, 610065, China}

\begin{abstract}
In this paper,  we study the Joule-Thomson expansion for RN-AdS black holes immersed in perfect fluid dark matter. Firstly,  the negative cosmological constant could be interpreted as thermodynamic pressure and its conjugate quantity with the volume gave us more physical insights into the black hole. We derive the thermodynamic definitions and study the critical behaviour of this black hole. Secondly, the explicit expression of Joule-Thomson coefficient is obtained from the basic formulas of enthalpy and temperature. Then, we obtain the isenthalpic curve in $T-P$ graph and demonstrate the cooling-heating region by the inversion curve. At last, we derive the ratio of minimum inversion temperature to critical temperature and the inversion curves in terms of charge $Q$ and parameter $\lambda$.  
\end{abstract}
\maketitle

\section{Introduction}
Since Bekenstein and Hawking's first study, black holes as thermodynamic systems have been an interesting research field in the theory of gravity \cite{Hawking:1974rv,Bardeen:1973gs,Bekenstein:1974ax,Hawking:1974sw,Hawking:1982dh}. The properties of AdS black hole thermodynamics have been studied since the early work of  Hawking-Page phase transition \cite{Hawking:1982dh}. Another notable thing is that the dark energy which corresponds to the cosmological constant is introduced to the first law of black hole thermodynamics. Particularly, the negative cosmological constant could be interpreted as thermodynamic pressure and treated as a thermodynamic variable \cite{Caldarelli:1999xj,Padmanabhan:2002sha}. 
And its conjugate quantity could be regarded as volume of black holes once the thermodynamic laws of black hole mechanics were generalized \cite{Kastor:2009wy}. Based on this idea, the charged AdS black hole thermodynamic properties and plenty of other studies were successfully investigated \cite{Liang:2020hjz,Saurabh:2020zqg,Cai:2001dz,Zou:2013owa,Hendi:2017fxp,Kubiznak:2016qmn,Kubiznak:2012wp,Altamirano:2014tva,Niu:2011tb,Myung:2008eb,Hendi:2015hoa,Toth:2011ab,Crisford:2017zpi,Hendi:2012um,Cai:2013qga,Wang:2018xdz}.

In the extended phase space (including $P$ and $V$ terms in the black hole thermodynamics), phase transition of charged AdS black holes is remarkably coincident with van der Waals liquid-gas phase transition \cite{Chamblin:1999tk,Chamblin:1999hg,Kubiznak:2012wp}. An interesting aspect of van der Waals system is Joule-Thomson expansion which indicates that the heating and cooling zone emerge through the throttling process. Since the phase structure and the critical behaviour of AdS black holes are similar to van der Waals system, the Joule-Thomson expansion of AdS black holes was firstly investigated by \"Okc\"u and Ayd\i{}ner \cite{Okcu:2016tgt}. In the extended thermodynamics, one identifies the enthalpy $H$ with the mass of the black hole \cite{Kastor:2009wy}. Joule-Thomson expansion is characterized by the invariance of enthalpy, so the throttling process is also an isenthalpic process. The slope of the isenthalpy curves equals the Joule-Thomson coefficient $\mu$ which determines the final change of temperature in this system. One can use the sign of the Joule-Thomson coefficient to determine the heating and cooling zone. Subsequently, there were many studies on Joule-Thomson expansion in various black holes \cite{Haldar:2018cks,Mo:2018qkt,Cisterna:2018jqg,Pu:2019bxf,Ghaffarnejad:2018exz,Lan:2018nnp,Feng:2020swq,Bi:2020vcg,Guo:2020qxy,Huang:2020xcs,Hegde:2020xlv,K.:2020rzl,Rostami:2019ivr,Nam:2019zyk,Yekta:2019wmt,Li:2019jcd,Rizwan:2018mpy,Chabab:2018zix,Mo:2018rgq,Okcu:2017qgo,Meng:2020csd,Nam:2020gud}. 

The standard model of cosmology suggests that our universe is compiled of dark matter, dark energy and baryonic matter, and the existence of dark matter and dark energy has been proven by several experiments and observations \cite{Ade:2015xua}. There exists a greater velocity than expected which means that more mass is required and it is thought to be provided by dark matter \cite{Das:2020boe,DeRisi:2012af}. As one of the dark matter candidates, the perfect fluid dark matter (PFDM) has been considered \cite{Rahaman:2010xs,Kiselev:2003ah}.  While astrophysical observations show that there exists a supermassive black hole surrounded by dark matter halo \cite{Akiyama:2019eap,Akiyama:2019cqa} and many black hole solutions within dark matter have been proposed. Particularly, spherically symmetric black hole solutions surrounded by PFDM have been obtained \cite{Kiselev:2004vy,Kiselev:2004py,Kiselev:2003ah,Li:2012zx}. 
Plenty of properties about this kind of black holes were discussed in \cite{Shaymatov:2020wtj,Das:2020yxw,Sadeghi:2020xtc,Hou:2018avu,Narzilloev:2020qtd,Ma:2020dhv,Zhang:2020mxi,Shaymatov:2020bso,Hendi:2020zyw,Rizwan:2018rgs,Haroon:2018ryd,Xu:2017bpz}. Here we intend to study the Joule-Thomson expansion of RN-AdS black hole immersed in PFDM.

This paper is organized as follows. In Section \eqref{PFDM}, we investigate the thermodynamic properties of RN-AdS black holes immersed in PFDM. In Section \eqref{JT}, we discuss the Joule-Thomson expansion of this kind of  black holes, which includes the Joule-Thomson coefficient, the inversion curves and the isenthalpic curves. Furthermore, we compare the critical temperature and the minimum of inversion temperature, and the influence of the PFDM parameter $\lambda$ and charge $Q$ on inversion curves is discussed. Finally, we discuss our result in Section \eqref{Conclusion}.

\section{RN-AdS black hole immersed in PFDM}\label{PFDM}
The minimum coupling of dark matter field with gravity, electromagnetic field and cosmological constant is described by the action \cite{Li:2012zx,Kiselev:2003ah,Xu:2016ylr},
\begin{equation}
\label{action}
S=\int \sqrt{-g}\left(\mathcal{L}_{\text{DM}}+\frac{F_{\mu \nu } F^{\mu \nu }}{4}-\frac{\Lambda }{8 \pi  G}+\frac{R}{16 \pi  G}\right)d^4x, 
\end{equation}
where $G$ is gravitational constant, $\Lambda$ is cosmological constant, $F_{\mu \nu }$ is electromagnetic tensor and $\mathcal{L}_{\text{DM}}$ is the Lagrangian which is related to the density of PFDM. 

The spacetime metric of static and spherically symmetric RN-AdS black holes immersed in PFDM is defined as \cite{Shaymatov:2020wtj,Xu:2016ylr}
\begin{equation}\label{ds^2}
ds^2 = -f (r) dt^2 + f (r)^{-1}dr^2 + r^2 d\Omega^2,
\end{equation}
where $d \Omega^{2}=d \theta^{2}+\sin ^{2} \theta d \phi^{2}$ and $f(r)$ is given by
 \begin{equation}\label{f(r)}
f(r)=1-\frac{2 M}{r}+\frac{Q^2}{r^2}-\frac{\Lambda}{3}r^2 +\frac{\lambda}{r}  \ln \left(\frac{r}{\left| \lambda \right| }\right),
\end{equation}
with $Q$ being charge of the black hole and $\lambda$ related to the dark matter density and pressure. This metric reduces to the RN-AdS black hole when $\lambda = 0$. For the given spacetime metric in the case of $\lambda \neq 0$ the stress energy-momentum tensor of the dark matter distribution is that of an anisotropic perfect fluid  reads as
\begin{equation}\label{tensor}
T_\nu^\mu=diag(-\rho,p_r,p_\theta,p_\phi),
\end{equation}
where density, radial and tangential pressures are given by \cite{Shaymatov:2020bso}
\begin{equation}\label{rho}
\rho=-p_r=\frac{\lambda}{8\pi r^3},  \quad p_\theta=p_\phi=\frac{\lambda}{16\pi r^3}.
\end{equation}
Note that for a PFDM distribution we shall restrict ourselves to the case $\lambda > 0$ which gives positive energy density. 

The cosmological constant $\Lambda$ is often parameterized by the AdS radius $l$ according to
\begin{equation}\label{L}
\Lambda=-\frac{(d-1)(d-2)}{2l^2}.
\end{equation}
And the negative cosmological constant is related to thermodynamic pressure $P$,
\begin{equation}
\label{P-L}
P=-\frac{\Lambda}{8\pi}.
\end{equation}
The entropy of the black hole at the horizon is given by
\begin{equation}\label{SS}
S= {\pi r_{+}^2},  
\end{equation}

The event horizon $r_{+}$ is the solution of $f(r_{+})=0$, which indicates the black hole mass $M$ can be expressed in event horizon $r_+$ as
\begin{equation}
\label{M}
M=\frac{r_{+}}{2}+\frac{4}{3} \pi P r_{+}^3+\frac{Q^2}{2 r_{+}}+\frac{1}{2} \lambda  \ln \left(\frac{r_{+}}{\lambda}\right).
\end{equation}
The first law of thermodynamics can be expressed as \cite{Xu:2016ylr}
\begin{equation}
dM=TdS+\Phi dQ+VdP+Ad\lambda.
\end{equation}
where the thermodynamic volume $V$ of black hole,  the electrostatic potential $\Phi$ and the conjugate quantity $A$ are given by
\begin{equation}\label{SV}
V=\frac{4 }{3}\pi  r_{+}^3, \indent \Phi=\frac{Q}{r_+}, \indent A=\frac{1}{2}\ln\left(\frac{r_{+}}{\lambda}\right).
\end{equation}
	
The Hawking temperature $T$ is obtained as
\begin{equation}\label{Tr}
T=\left(\frac{\partial M}{\partial S}\right)_{Q,P,\lambda} =\frac{\lambda }{4 \pi r_{+}^2}+2 P r_{+}+\frac{1}{4 \pi r_{+}}-\frac{Q^2}{4 \pi r_+^3}.
\end{equation}

\begin{figure}[H]\centering
 	\begin{subfigure}[b]{.45\linewidth}
	\centering
	\includegraphics[width=.9\linewidth]{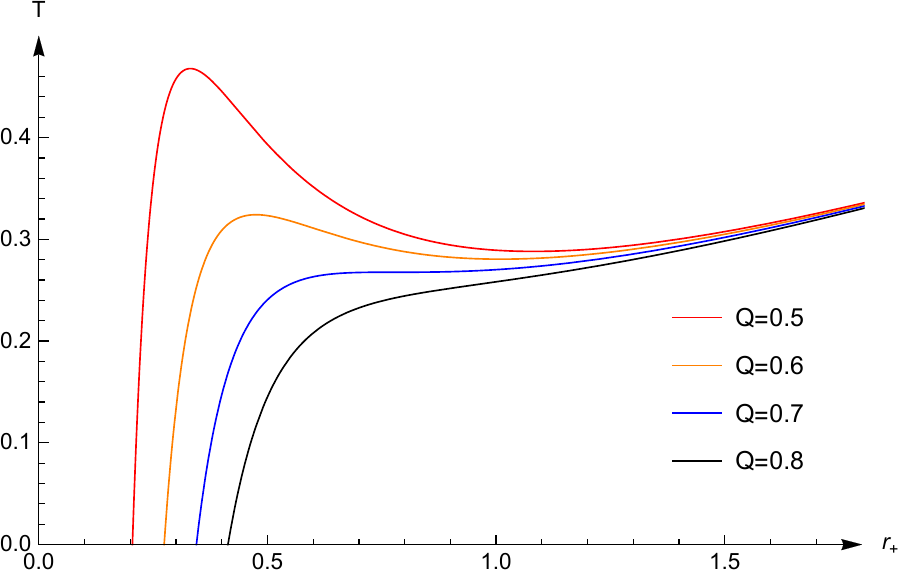}
	\end{subfigure}
\begin{subfigure}[b]{.45\linewidth}
	\centering
	\includegraphics[width=0.9\linewidth]{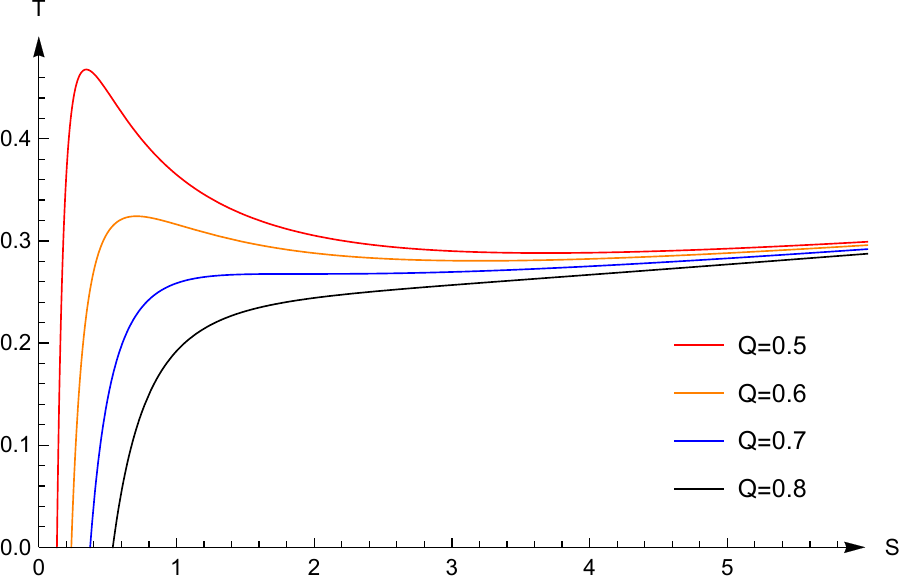}
\end{subfigure}
\caption{
The temperature $T$ of RN-AdS black hole immersed in PFDM versus $r_+$ and $S$ for different values of $Q$, where $\lambda=1$ and $P =0.075$. }
\label{fig:Tr}
\end{figure}

The Hawking temperature $T$ versus event horizon $r_+$ and entropy $S$ are shown in FIG.(\ref{fig:Tr}) for different values of $Q$ with $\lambda=1$ and $P =0.075$. The Hawking temperature is zero corresponding to the extreme black hole. Since the slope of the $T-S$ graph is related to specific heat capacity, it’s positive or negative values  determine the stability of the system with respect to fluctuations. Hence the FIG. (\ref{fig:Tr}) shows that there exists a critical point which indicates the phase transition \cite{Rizwan:2018mpy}. In this case, we think that the asymptotic AdS black hole in a charge fixed canonical ensemble shows a first-order phase transition similar to that of a van der Waals fluids.

Next we will study the critical behaviour of this black hole. From Eq. (\ref{Tr}), we get the expression of pressure,
\begin{equation} \label{Pr}
P(r_{+},T)=\frac{T}{2  r_{+}}-\frac{1}{8 \pi  {r_{+}^2}}+\frac{Q^2}{8 \pi  {r_{+}^4}}-\frac{\lambda }{8 \pi  {r_{+}^3}}.
\end{equation}
At the critical point, we have
\begin{equation}\label{partial}
\left.\frac{\partial P}{\partial r_{+}}\right|_{r=r_{c}}=\left.\frac{\partial^{2} P}{\partial r_{+}^{2}}\right|_{r=r_{c}}=0,
\end{equation}
\begin{equation}\label{partialT}
	\left.\frac{\partial T}{\partial r_{+}}\right|_{r=r_{c}}=\left.\frac{\partial^{2} T}{\partial r_{+}^{2}}\right|_{r=r_{c}}=0,
\end{equation}
so the critical point $r_{c}$ is obtained as
\begin{equation}\label{rc}
r_c=\frac{1}{2} \sqrt{9 \lambda ^2+24 Q^2}-\frac{3 \lambda }{2}.
\end{equation}

By using the Eqs. (\ref{Tr}), (\ref{Pr}) and (\ref{rc}), we can determine the critical temperature $T_c$ and critical pressure $P_c$,
\begin{align}\label{TcL}
T_c&=\frac{16 Q^2-3 \lambda  \left(\sqrt{9 \lambda ^2+24 Q^2}-3 \lambda \right)}{\pi  \left(\sqrt{9 \lambda ^2+24 Q^2}-3 \lambda \right)^3}, \\
P_c&=\frac{\lambda  \left(3 \lambda -\sqrt{9 \lambda ^2+24 Q^2}\right)+6 Q^2}{\pi  \left(\sqrt{9 \lambda ^2+24 Q^2}-3 \lambda \right)^4}.
\end{align}

\begin{figure}[H]\centering
 	\begin{subfigure}[b]{.4\linewidth}
	\centering
	\includegraphics[width=0.95\linewidth]{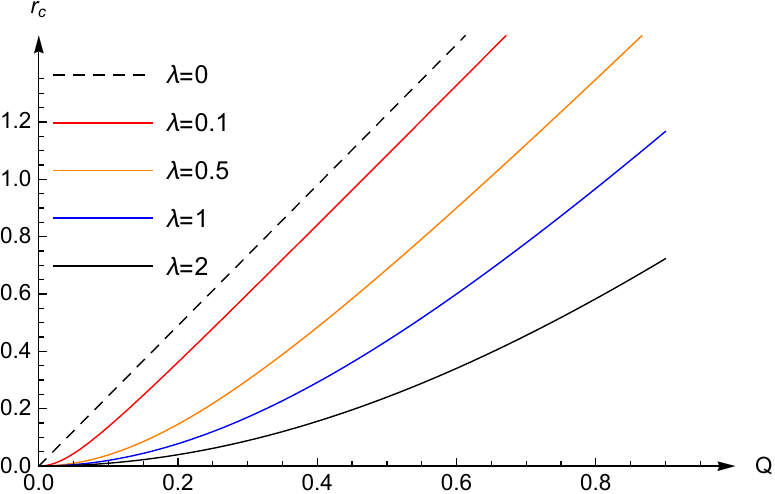}	
	\label{fig:rc}
	\end{subfigure}
\begin{subfigure}[b]{.4\linewidth}
	\centering
	\includegraphics[width=.95\linewidth]{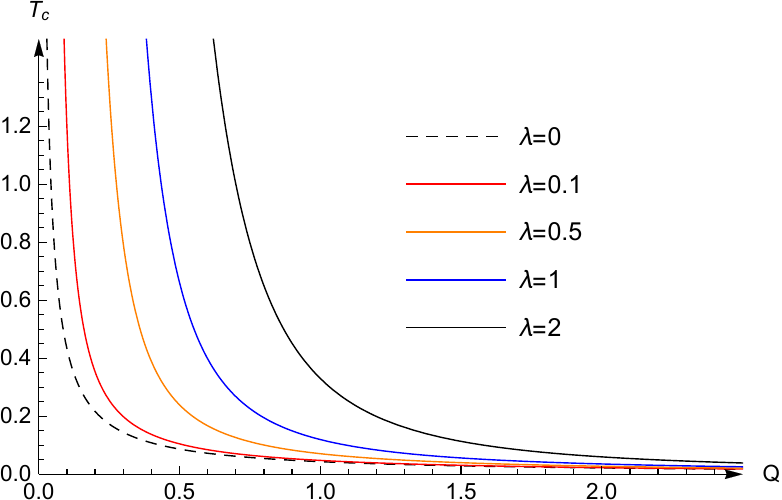}
	\label{fig:Tc}
\end{subfigure}
\caption{Critical radius $r_c$ and critical temperature $T_c$ versus $Q$ of RN-AdS black hole immersed in PFDM.}
\label{rcTc}
\end{figure}

It is clear that the critical variables depend on the charge $Q$ and parameters $\lambda$. Here we focus on the critical temperature because it will be useful to analyze the Joule-Thomson expansion of the black hole in the next section. One can plot the critical point $r_c$ and the critical temperature $T_c$ with varying $Q$ and $\lambda$ in FIG. (\ref{rcTc}). The critical temperature decreases while charge $Q$ increases. If the temperature is less than the critical temperature $T_c$, the black hole can undergo a first-order phase transition between the small black hole and the large black hole, which is similar to the van der Waals fluids.  Furthermore, one can go back to the RN-AdS case as in \cite{Okcu:2016tgt} by taking the limit $\lambda \rightarrow 0$. Thus, the influence of PFDM on the critical behaviours of black hole is obvious.
We derive the thermodynamic definitions of RN-AdS black holes immersed in PFDM, and we study the critical behaviour of this black hole. In the next section, we investigate the Joule-Thomson expansion of RN-AdS black holes immersed in PFDM.

\section{Joule-Thomson Expansion}\label{JT}
 The instability of the black holes appears in the process of Joule-Thomson expansion. And we introduce Joule-Thomson coefficient $\mu$ to determines the cooling and heating phases of the isenthalpic expansion. 

The specific heat at constant pressure for the black holes is obtained according to the first law of thermodynamics,
\begin{equation}\label{cp}
C_p=T \left(\frac{\partial S}{\partial T}\right)_{P,Q,\lambda}.
\end{equation}
Regarding parameters $Q, P, \lambda$ as constant, and substituting the Eqs. (\ref{SV}) and (\ref{Tr}) into (\ref{cp}), we have the isobaric heat capacity of this black hole,
\begin{equation}
C_p=\frac{2 \pi  r_{+}^2 \left[r_{+} \left(\lambda +8 \pi  P r_{+}^3+r_{+}\right)-Q^2\right]}{r_{+} \left(-2 \lambda +8 \pi  P r_{+}^3-r_{+}\right)+3 Q^2}.
\label{Cp}
\end{equation}
For the Joule-Thomson expansion of the van der Waals fluids, the fluids passes through a porous plug from one side to the other with pressure declining during throttling process. Therefore, we apply this corresponding concept to black hole thermodynamics and the enthalpy $M$ of the black hole keeps constant throughout this process. Besides, the partial
differential of temperature versus pressure of the black hole is defined as Joule-Thomson
coefficient $\mu$,
\begin{equation}
\mu=\left(\frac{\partial T}{\partial P}\right)_{M} =\frac{1}{C_P}\left [T\left( \frac{\partial V}{\partial T}\right)_P-V\right]
\label{u}.
\end{equation}
By substituting Eqs (\ref{Tr}), (\ref{Cp}) and (\ref{SV}) into (\ref{u}), we obtain
\begin{equation}
\mu =\frac{2 r_{+} \left[r_{+} \left(5 \lambda +16 \pi  P r_{+}^3+4 r_{+}\right)-6 Q^2\right]}{3 \left[r_{+} \left(\lambda +8 \pi  P r_{+}^3+r_{+}\right)-Q^2\right]}
\label{u1}.
\end{equation}

\begin{figure}[H]\centering
	\includegraphics[width=0.5\linewidth]{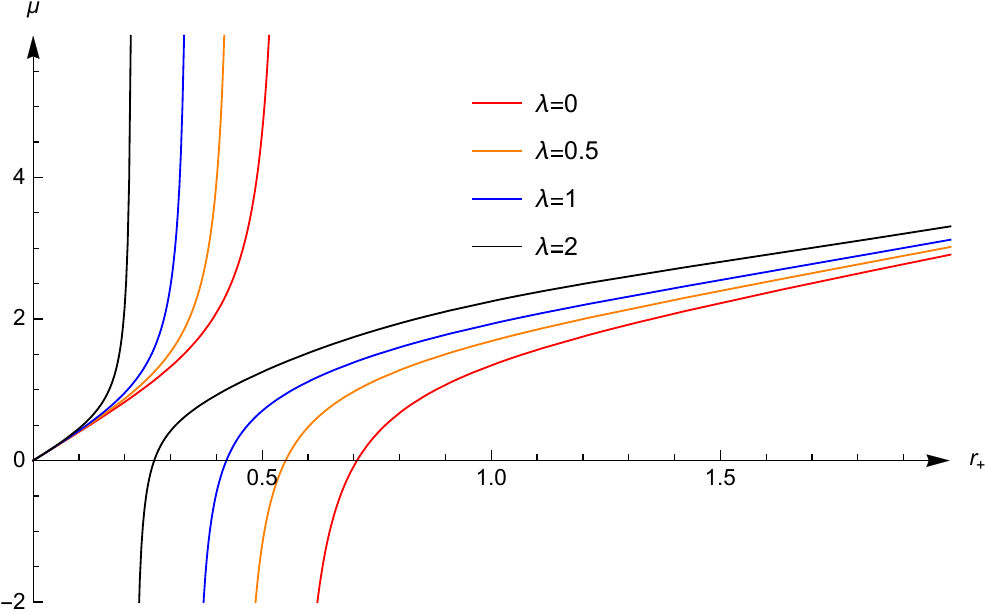}
	\caption{Joule-Thomson coefficient $\mu$ of RN-AdS black hole immersed in PDFM for $Q =0.7$, $P=0.075$ and $\lambda = 0,0.5,1$ and $2$.}
	\label{fig:mu}
\end{figure}

The Joule-Thomson coefficient $\mu$ versus the horizon $r_+$ is shown in FIG. (\ref{fig:mu}) where the parameter $\lambda = 0,0.5,1$ and $2$, pressure $P = 0.075$ and charges $Q=0.7$. There exist both a divergence point and a zero point  at each different value of $\lambda$ and the influence of PFDM on JT coefficient is apparent. The divergence point here reveals the information of Hawking temperature and corresponds to the extremal black hole, and it is clear that the divergence point of the blue curve in FIG. (\ref{fig:mu}) is consistent with the zero point of Hawking temperature in FIG. (\ref{fig:Tr}) in the case of $Q=0.7$.
 Therefore, we are only interested in the right part of FIG. (\ref{fig:mu}) where the temperature is above zero. Besides, the Joule-Thomson coefficient of RN-AdS black hole immersed in PFDM can be both positive and negative, which means that there remain both cooling and heating stages. For $\mu>0$, the temperature of the black hole decreases in other words the black hole is
cooling down when pressure goes down in the process of throttling. On the contrary, $\mu<0$ means that the black hole heats up after the throttling process. So there are inversion points at which the cooling-heating transition occurring when $\mu=0$, and the temperature of the black hole at $\mu=0$ is called the inversion temperature.

\begin{figure}[H]\centering
\begin{subfigure}	
	{.45\linewidth}
	\centering
	\includegraphics[width=0.9\linewidth]{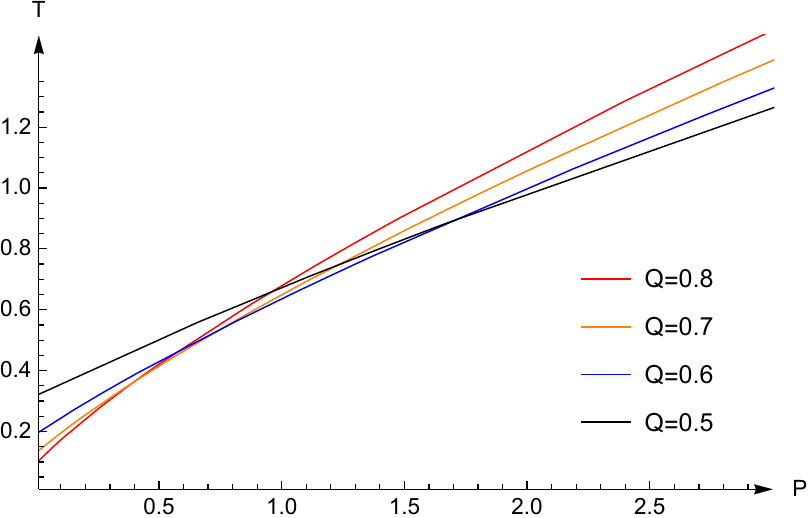}
	\caption{ $\lambda= 1$, and $Q = 0.5, 0.6, 0.7, 0.8$.}
\end{subfigure}
\begin{subfigure}
{0.45\linewidth}
	\centering
	\includegraphics[width=0.9\linewidth]{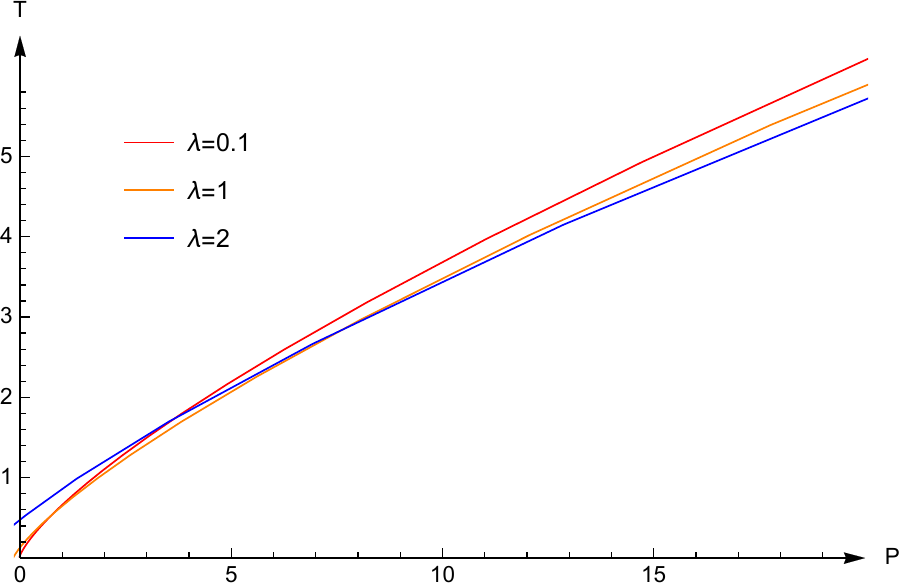}
		\caption{$Q= 0.7$ and $\lambda= 0.1, 1, 2$.}	
\end{subfigure}
	\caption{Inversion curves of RN-AdS black hole immersed in PFDM.}
	\label{fig:tp}
\end{figure}

One can obtain the inversion temperature $T_i=V \displaystyle\frac{\partial T}{\partial V}$  by $\mu=0$. While using the Eqs. (\ref{SV}) and (\ref{Tr}), inversion temperature $T_i$ in terms of inversion pressure $P_i$ is expressed as
\begin{equation}
T_i=V \left(\frac{\partial T}{\partial V}\right)_P=\frac{2 P_i r_{+}}{3}+\frac{Q^2}{4 \pi  r_{+}^3}-\frac{\lambda }{6 \pi  r_{+}^2}-\frac{1}{12 \pi  r_{+}}. \label{Ti}
\end{equation}
Bringing this equation into Eq.(\ref{Pr}) at $P = P_i$, we derive
\begin{align}
P_i&=\frac{3 Q^2}{8 \pi  r_+^4}-\frac{5 \lambda }{16 \pi  r_+^3}-\frac{1}{4 \pi  r_+^2} \label{Pi}, \\
T_i&=\frac{Q^2}{2 \pi  r_+^3}-\frac{3 \lambda }{8 \pi  r_+^2}-\frac{1}{12 \pi  r_+^2}-\frac{1}{6 \pi  r_+}.
\end{align}
From above results, the inversion curves $T_i-P_i$ in FIG. (\ref{fig:tp}) with different $Q$ and $\lambda$ is plotted, which is similar to the case of charged RN AdS black hole \cite{Okcu:2016tgt}. At low pressure, the inversion temperature $T_i$ decreases with the increase of charge $Q$ and increases with the increase of $\lambda$. This phenomenon is just the opposite of high pressure situation. In addition, there are some numerical effects of PFDM on the inversion temperature, but no giant differences from charged AdS black hole. It can be observed that the inversion temperature still increases monotonically with the growth of inversion pressure, and the inversion curves are not closed which is different from the situation of the van der Waals fluids. 

\begin{figure}[H]
\centering
	\begin{subfigure}{0.4\linewidth}
		\centering
		\includegraphics[width=0.9\linewidth]{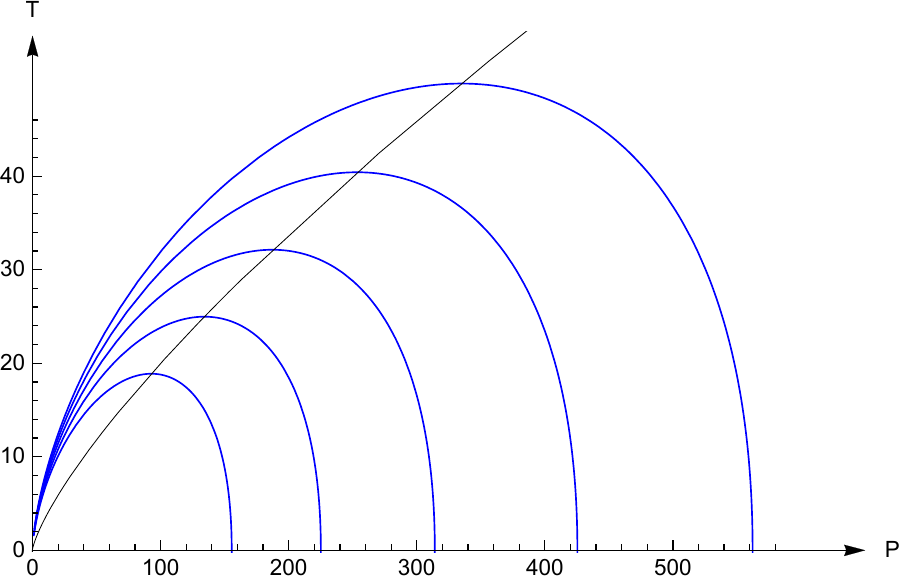}
		\caption{$\lambda = 1$, $Q=0.7$ and $M=2, 2.25, \\2.5, 2.75, 3$.}
		\label{fig:tpblueq0}
	\end{subfigure}
	\begin{subfigure}{0.4\linewidth}
		\centering
		\includegraphics[width=0.9\linewidth]{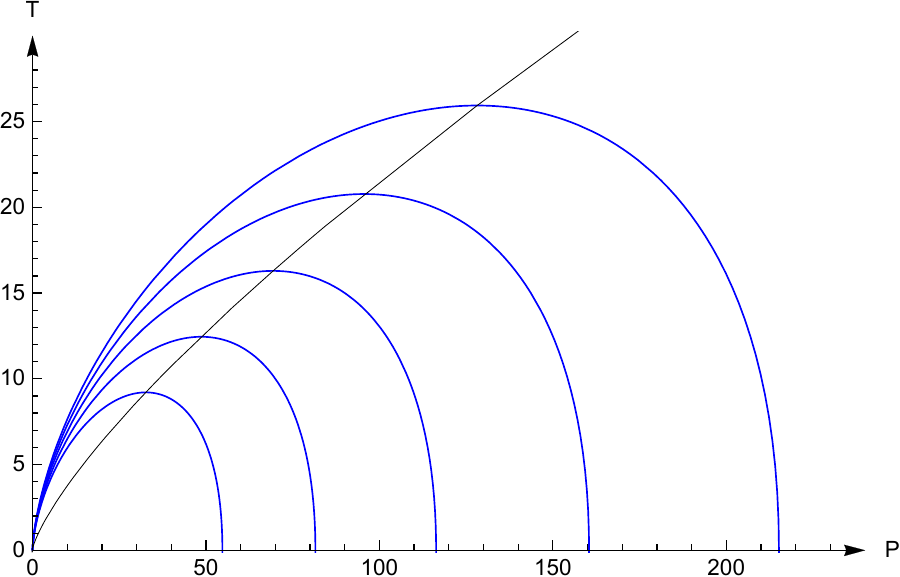}
		\caption{$\lambda = 1$, $Q=0.8$ and $M=2, 2.25, \\2.5, 2.75, 3$.}
		\label{fig:Red}
	\end{subfigure}
	\begin{subfigure}{0.4\linewidth}
		\centering
		\includegraphics[width=0.9\linewidth]{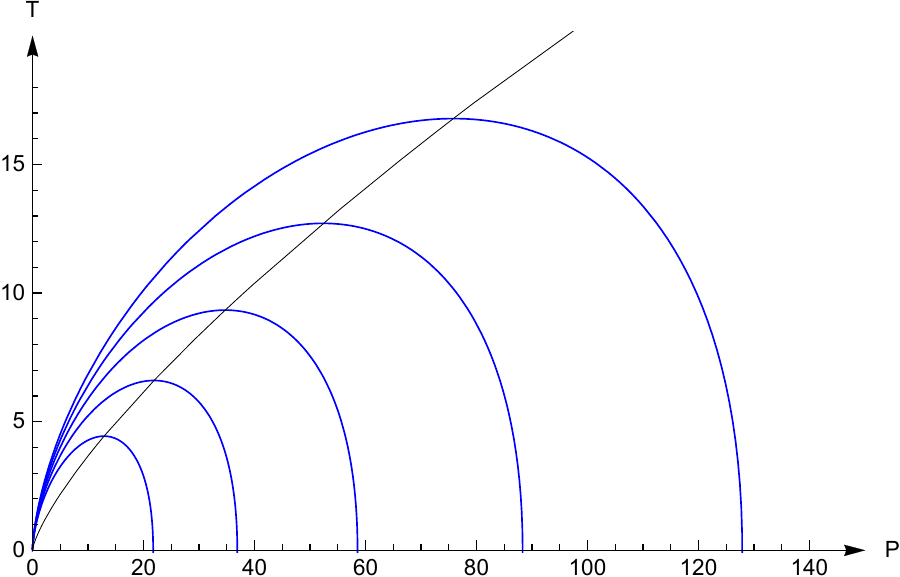}
		\caption{$\lambda = 1$, $Q=0.7$, and $M=2,2.25,\\ 2.5, 2.75, 3$.}
		\label{fig:Orange}
	\end{subfigure}
	\begin{subfigure}{0.4\linewidth}
		\centering
		\includegraphics[width=0.9\linewidth]{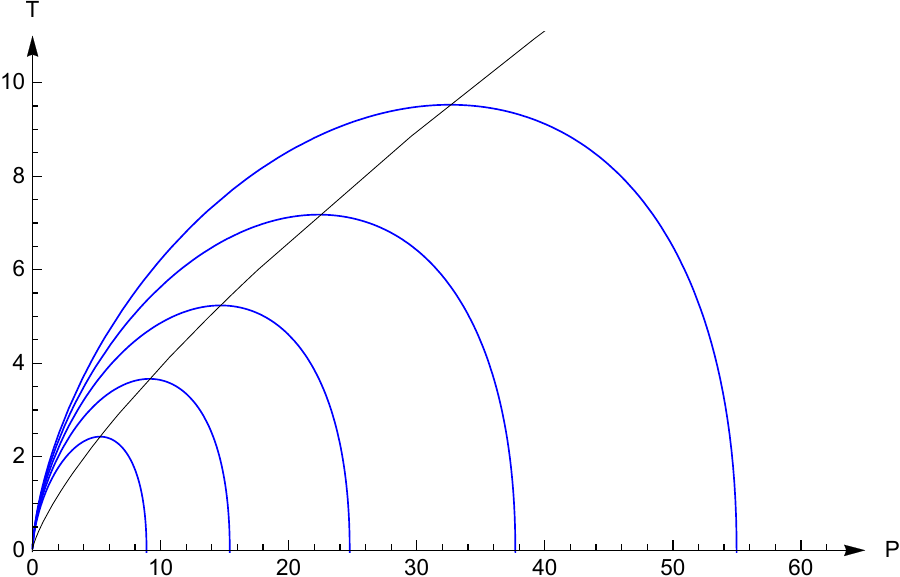}
		\caption{$\lambda =0.1,$ $Q=0.8$ and $M=2, 2.25, \\2.5, 2.75, 3$.}
		\label{fig:green}
	\end{subfigure}
	\caption{The isenthalpic curves of RN-AdS black hole immersed in PFDM. From bottom to top, the isenthalpic curves correspond to increasing values of $M$.}
	\label{H}
\end{figure}

The Joule-Thomson expansion occurs in the isenthalpic process. And for a black hole, enthalpy is mass $M$. The isenthalpic curves can be obtained from Eqs. (\ref{M}) and (\ref{Pr}) in FIG. (\ref{H}) where inversion curves and isenthalpic curves are both presented.
The intersections of the black inversion curves and the blue isenthalpic curves indicating the cooling-heating transition coincide with the extreme points of isenthalpic curves. And the isenthalpic curves have positive slope above the inversion curves indicating that the cooling occurs. And the isenthalpic curves have negative slope under the inversion curves indicating that the heating occurs. We can conclude that the region above this inversion curve is cooling zone while the one below is heating zone. The temperature rises when the pressure goes down in the heating zone. In contrast, the temperature is lower with reducing pressure in the cooling zone. Moreover, comparing the four graphs the temperature and pressure decrease with larger charge $Q$ or smaller $\lambda$, and PFDM statistically have an influence on the isenthalpic curves.  

\begin{figure}[H]
\centering
\label{key}\includegraphics[width=0.5\linewidth]{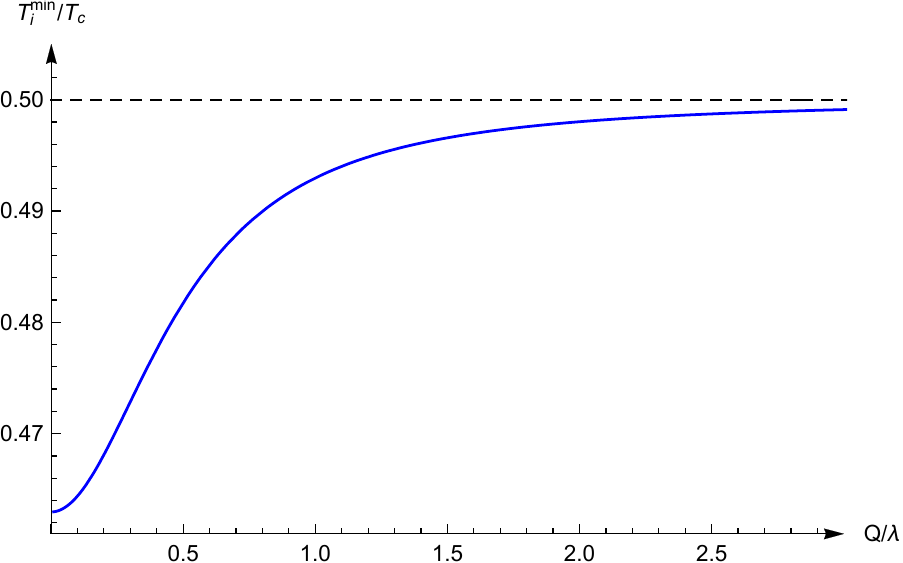}
	\caption{The ratio ${T_i^{\min }}/{T_c}$ of RN-AdS black hole immersed in PFDM.}
	\label{fig:tmintc}
\end{figure}

Furthermore, one can obtain the minimum inversion temperature $T_i^{\min }$ corresponding to $P_i = 0$ as
\begin{equation}
T_i^{\min }=\frac{4 \left[\lambda  \left(5 \lambda -\sqrt{25 \lambda ^2+96 Q^2}\right)+16 Q^2\right]}{\pi  \left(\sqrt{25 \lambda ^2+96 Q^2}-5 \lambda \right)^3}.
\label{Timin}
\end{equation}
The ratio of inversion temperature to critical temperature ${T_i^{\min }}/{T_c}$ is obtained according to Eqs. (\ref{Timin}) and (\ref{TcL}), 
\begin{equation}
\frac{T_i^{\min }}{T_c}=\frac{4 \left(\sqrt{9 \lambda ^2+24 Q^2}-3 \lambda \right)^3 \left[\lambda  \left(5 \lambda -\sqrt{25 \lambda ^2+96 Q^2}\right)+16 Q^2\right]}{\left(\sqrt{25 \lambda ^2+96 Q^2}-5 \lambda \right)^3 \left[16 Q^2-3 \lambda  \left(\sqrt{9 \lambda ^2+24 Q^2}-3 \lambda \right)\right]}.
\end{equation}
 It has been shown in \cite{Okcu:2016tgt} that the ratio for the RN-AdS black holes equals $1/2$, but in the case of PFDM this value is corrected. In fact, the ratio grows with increasing $Q/\lambda$, and the ratio versus $Q/\lambda$ is shown in FIG. (\ref{fig:tmintc}).  When $Q/\lambda\rightarrow 0$, in other words $Q\rightarrow0$ or $\lambda \rightarrow \infty$, the ratio has a leading term which is 25/54,
\begin{equation}
\frac{T_i^{\min }}{T_c}=\frac{25}{54}+\frac{4 Q^2}{27 \lambda ^2}+O\left[ (Q/\lambda)^4\right],
\end{equation}
and it is close to the situation of Schwarzschild black hole within PFDM. 

For $Q/\lambda\rightarrow\infty$, 
\begin{equation}
\frac{T_i^{\min }}{T_c}=\frac{1}{2}-\frac{\lambda ^2}{128 Q^2}+O\left[ (\lambda/Q) ^3\right].
\end{equation}

Furthermore, by taking the limit  $\lambda \rightarrow0 $, the ratio is approaching $1/2$ corresponding to the RN-AdS case \cite{Okcu:2016tgt}.

\section{Conclusion}\label{Conclusion}
In this paper,  we discuss the Joule-Thomson expansion for RN-AdS black holes immersed in PFDM. In the extended phase space, the cosmological constant is identified with the pressure and its conjugate quantity as the thermodynamic volume. Since the black hole mass is interpreted as enthalpy,  an isenthalpic process can be applied to calculate the temperature change during the Joule-Thomson expansion. By computing the Joule-Thomson coefficient $\mu$, the divergent point coincides with the extremal black hole while the zero point is the inversion point that determines the transition from heating/cooling phases. And we calculate the inversion curves in the $T-P$ plane as well as the corresponding isenthalpic curves. . And we can use inversion curve distinguish the cooling and heating regions for different values of $\lambda$ and $Q$. Our calculations show that the effect of PFDM density parameter $\lambda$ on Joule-Thomson coefficient and the inversion temperature is obvious. In addition, one can easily go back to the RN-AdS case by taking the limit $\lambda \rightarrow 0$. We also discuss the isenthalpic curves with different values of $M$ during the throttling process.  

Furthermore, we derive the ratio of minimum inversion temperature to critical temperature and the inversion curves in terms of charge $Q$ and parameter $\lambda$. It has been shown in \cite{Okcu:2016tgt} that the ratio for the charged RN-AdS black holes is equal to $1/2$, but in the case of PFDM the ratio depends on the charge $Q$ and parameter $\lambda$.  In fact, the ratio of ${T_i^{\min }}/{T_c}$ depends on $Q/\lambda$, which is shown in FIG. (\ref{fig:tmintc}). The ratio is about 25/54 for small $Q/\lambda$, and it is corresponding to the situation of RN-AdS black holes for $\lambda=0$.

\begin{acknowledgments}
We are grateful to thank Shihao Bi and Feiyu Yao for useful discussions. This work is supported by NSFC (Grant No.11947408 and 11875196).
\end{acknowledgments}


\begin{thebibliography}{99}
\bibitem{Hawking:1974rv}
S.~W.~Hawking,
Black hole explosions,
Nature \textbf{248},(1974) 30-31. 

\bibitem{Bardeen:1973gs}	
J.~M.~Bardeen, B.~Carter and S.~W.~Hawking,
The Four laws of black hole mechanics,
Commun. Math. Phys. \textbf{31} (1973), 161-170.

\bibitem{Hawking:1974sw}
S.~W.~Hawking,
Particle Creation by Black Holes, 
Commun. Math. Phys. \textbf{43} (1975), 199-220.

\bibitem{Bekenstein:1974ax}
J.~D.~Bekenstein,
Generalized second law of thermodynamics in black hole physics,
Phys. Rev. D \textbf{9} (1974), 3292-3300.
	
\bibitem{Hawking:1982dh}
S.~W.~Hawking and D.~N.~Page,
Thermodynamics of Black Holes in anti-De Sitter Space,
Commun. Math. Phys. \textbf{87} (1983), 577.

\bibitem{Caldarelli:1999xj}
M.~M.~Caldarelli, G.~Cognola and D.~Klemm,
Thermodynamics of Kerr-Newman-AdS black holes and conformal field theories,
Class. Quant. Grav. \textbf{17} (2000), 399-420

\bibitem{Padmanabhan:2002sha}
T.~Padmanabhan,
Classical and quantum thermodynamics of horizons in spherically symmetric space-times,
Class. Quant. Grav. \textbf{19} (2002), 5387-5408

\bibitem{Kastor:2009wy}
D.~Kastor, S.~Ray and J.~Traschen,
Enthalpy and the Mechanics of AdS Black Holes,
Class. Quant. Grav. \textbf{26} (2009), 195011.

\bibitem{Kubiznak:2016qmn}
D.~Kubiznak, R.~B.~Mann and M.~Teo,
Black hole chemistry: thermodynamics with Lambda,
Class. Quant. Grav. \textbf{34} (2017) no.6, 063001.

\bibitem{Cai:2001dz}
R.~G.~Cai,
Gauss-Bonnet black holes in AdS spaces,
Phys. Rev. D \textbf{65} (2002), 084014.

\bibitem{Zou:2013owa}
D.~C.~Zou, S.~J.~Zhang and B.~Wang,
Critical behavior of Born-Infeld AdS black holes in the extended phase space thermodynamics,
Phys. Rev. D \textbf{89} (2014) no.4, 044002.
 
\bibitem{Hendi:2017fxp}
S.~H.~Hendi, R.~B.~Mann, S.~Panahiyan and B.~Eslam Panah,
Van der Waals like behavior of topological AdS black holes in massive gravity,
Phys. Rev. D \textbf{95} (2017) no.2, 021501.

\bibitem{Altamirano:2014tva}
N.~Altamirano, D.~Kubiznak, R.~B.~Mann and Z.~Sherkatghanad,
Thermodynamics of rotating black holes and black rings: phase transitions and thermodynamic volume,
Galaxies \textbf{2} (2014), 89-159.

\bibitem{Niu:2011tb}
C.~Niu, Y.~Tian and X.~N.~Wu,
Critical Phenomena and Thermodynamic Geometry of RN-AdS Black Holes,
Phys. Rev. D \textbf{85} (2012), 024017.

\bibitem{Myung:2008eb}
Y.~S.~Myung, Y.~W.~Kim and Y.~J.~Park,
Thermodynamics and phase transitions in the Born-Infeld-anti-de Sitter black holes,
Phys. Rev. D \textbf{78} (2008), 084002.

\bibitem{Hendi:2015hoa}
S.~H.~Hendi, B.~Eslam Panah and S.~Panahiyan,
Einstein-Born-Infeld-Massive Gravity: adS-Black Hole Solutions and their Thermodynamical properties,
JHEP \textbf{11} (2015), 157.


\bibitem{Saurabh:2020zqg}
Saurabh and K.~Jusufi,
Imprints of Dark Matter on Black Hole Shadows using Spherical Accretions,
arXiv:2009.10599 [gr-qc].

\bibitem{Kubiznak:2012wp}
D.~Kubiznak and R.~B.~Mann,
P-V criticality of charged AdS black holes,
JHEP \textbf{07} (2012), 033.

\bibitem{Cai:2013qga}
R.~G.~Cai, L.~M.~Cao, L.~Li and R.~Q.~Yang,
P-V criticality in the extended phase space of Gauss-Bonnet black holes in AdS space,
JHEP \textbf{09} (2013), 005.


\bibitem{Hendi:2012um}
S.~H.~Hendi and M.~H.~Vahidinia,
Extended phase space thermodynamics and P-V criticality of black holes with a nonlinear source,
Phys. Rev. D \textbf{88} (2013) no.8, 084045.

\bibitem{Wang:2018xdz}
P.~Wang, H.~Wu and H.~Yang,
Thermodynamics and Phase Transitions of Nonlinear Electrodynamics Black Holes in an Extended Phase Space,
JCAP \textbf{04} (2019), 052.

\bibitem{Toth:2011ab}
G.~Z.~Toth,
Test of the weak cosmic censorship conjecture with a charged scalar field and dyonic Kerr-Newman black holes,
Gen. Rel. Grav. \textbf{44} (2012), 2019-2035.

\bibitem{Crisford:2017zpi}
T.~Crisford and J.~E.~Santos,
Violating the Weak Cosmic Censorship Conjecture in Four-Dimensional Anti\textendash{}de Sitter Space,
Phys. Rev. Lett. \textbf{118} (2017) no.18, 181101.

\bibitem{Liang:2020hjz}
J.~Liang, X.~Guo, D.~Chen and B.~Mu,
Remarks on the weak cosmic censorship conjecture of RN-AdS black holes with cloud of strings and quintessence under the scalar field,
arXiv:2008.08327 [gr-qc].


\bibitem{Chamblin:1999tk}
A.~Chamblin, R.~Emparan, C.~V.~Johnson and R.~C.~Myers,
Charged AdS black holes and catastrophic holography,
Phys. Rev. D \textbf{60} (1999), 064018.


\bibitem{Chamblin:1999hg}
A.~Chamblin, R.~Emparan, C.~V.~Johnson and R.~C.~Myers,
Holography, thermodynamics and fluctuations of charged AdS black holes,
Phys. Rev. D \textbf{60} (1999), 104026.

\bibitem{Okcu:2016tgt}
\"O.~\"Okc\"u and E.~Ayd\i{}ner,
Joule\textendash{}Thomson expansion of the charged AdS black holes,
Eur. Phys. J. C \textbf{77} (2017) no.1, 24.

\bibitem{Bi:2020vcg}
S.~Bi, M.~Du, J.~Tao and F.~Yao,
Joule-Thomson Expansion of Born-Infeld AdS Black Holes,
[arXiv:2006.08920 [gr-qc]].

\bibitem{Guo:2020qxy}
S.~Guo, Y.~Han and G.~P.~Li,
Joule–Thomson expansion of a specific black hole in f(R) gravity coupled with Yang–Mills field,
Class. Quant. Grav. \textbf{37} (2020) no.8, 085016.

\bibitem{Huang:2020xcs}
Y.~l.~Huang and S.~Guo,
Thermodynamic of the charged accelerating AdS black hole: P-V critical and Joule-Thomson expansion,
arXiv:2009.09401 [hep-th].

\bibitem{Hegde:2020xlv}
K.~Hegde, A.~Naveena Kumara, C.~L.~A.~Rizwan, A.~K.~M. and M.~S.~Ali,
Thermodynamics, Phase Transition and Joule Thomson Expansion of novel 4-D Gauss Bonnet AdS Black Hole,
arXiv:2003.08778 [gr-qc].

\bibitem{K.:2020rzl}
R.~K., C.~L.~A.~Rizwan, A.~Naveena Kumara, D.~Vaid and M.~S.~Ali,
Joule-Thomson Expansion of Regular Bardeen AdS Black Hole Surrounded by Static Anisotropic Quintessence Field,
arXiv:2002.03634 [gr-qc].

\bibitem{Rostami:2019ivr}
M.~Rostami, J.~Sadeghi, S.~Miraboutalebi, A.~A.~Masoudi and B.~Pourhassan,
Charged accelerating AdS black hole of $f(R)$ gravity and the Joule-Thomson expansion,
Int. J. Geom. Meth. Mod. Phys. \textbf{17} (2020) no.09, 2050136.

\bibitem{Nam:2019zyk}
C.~H.~Nam,
Heat engine efficiency and Joule-Thomson expansion of non-linear charged AdS black hole in massive gravity,
arXiv:1906.05557 [gr-qc].

\bibitem{Yekta:2019wmt}
D.~Mahdavian Yekta, A.~Hadikhani and \"O.~\"Okc\"u,
Joule-Thomson expansion of charged AdS black holes in Rainbow gravity,
Phys. Lett. B \textbf{795} (2019), 521-527.

\bibitem{Lan:2018nnp}
S.~Q.~Lan,
Joule-Thomson expansion of charged Gauss-Bonnet black holes in AdS space,
Phys. Rev. D \textbf{98} (2018) no.8, 084014.

\bibitem{Ghaffarnejad:2018exz}
H.~Ghaffarnejad, E.~Yaraie and M.~Farsam,
Quintessence Reissner Nordstr\"om Anti de Sitter Black Holes and Joule Thomson effect,
Int. J. Theor. Phys. \textbf{57} (2018) no.6, 1671-1682.

\bibitem{Pu:2019bxf}
J.~Pu, S.~Guo, Q.~Q.~Jiang and X.~T.~Zu,
Joule-Thomson expansion of the regular(Bardeen)-AdS black hole,
Chin. Phys. C \textbf{44} (2020) no.3, 035102.

\bibitem{Mo:2018qkt}
J.~X.~Mo and G.~Q.~Li,
Effects of Lovelock gravity on the Joule\textendash{}Thomson expansion,
Class. Quant. Grav. \textbf{37} (2020) no.4, 045009.

\bibitem{Cisterna:2018jqg}
A.~Cisterna, S.~Q.~Hu and X.~M.~Kuang,
Joule-Thomson expansion in AdS black holes with momentum relaxation,
Phys. Lett. B \textbf{797} (2019), 134883.

\bibitem{Li:2019jcd}
C.~Li, P.~He, P.~Li and J.~B.~Deng,
Joule-Thomson expansion of the Bardeen-AdS black holes,
Gen. Rel. Grav. \textbf{52} (2020) no.5, 50.

\bibitem{Rizwan:2018mpy}
A.~Rizwan C.L., N.~Kumara A., D.~Vaid and K.~M.~Ajith,
Joule-Thomson expansion in AdS black hole with a global monopole,
Int. J. Mod. Phys. A \textbf{33} (2019) no.35, 1850210.

\bibitem{Chabab:2018zix}
M.~Chabab, H.~El Moumni, S.~Iraoui, K.~Masmar and S.~Zhizeh,
Joule-Thomson Expansion of RN-AdS Black Holes in $f(R)$ gravity,
LHEP \textbf{02} (2018), 05.

\bibitem{Mo:2018rgq}
J.~X.~Mo, G.~Q.~Li, S.~Q.~Lan and X.~B.~Xu, Joule-Thomson expansion of $d$-dimensional charged AdS black holes,
Phys. Rev. D \textbf{98} (2018) no.12, 124032.

\bibitem{Okcu:2017qgo}
\"O.~\"Okc\"u and E.~Ayd\i{}ner,
Joule\textendash{}Thomson expansion of Kerr\textendash{}AdS black holes,
Eur. Phys. J. C \textbf{78} (2018) no.2, 123.

\bibitem{Haldar:2018cks}
A.~Haldar and R.~Biswas,
Joule-Thomson expansion of five-dimensional Einstein-Maxwell-Gauss-Bonnet-AdS black holes,
EPL \textbf{123} (2018) no.4, 40005.

\bibitem{Feng:2020swq}
Z.~W.~Feng, X.~Zhou and S.~Q.~Zhou,
Joule-Thomson expansion of higher dimensional nonlinearly charged AdS black hole in Einstein-PMI gravity,
arXiv:2009.02172 [gr-qc].

\bibitem{Nam:2020gud}
C.~H.~Nam,
Effect of massive gravity on Joule\textendash{}Thomson expansion of the charged AdS black hole,
Eur. Phys. J. Plus \textbf{135} (2020) no.2, 259.

\bibitem{Meng:2020csd}
Y.~Meng, J.~Pu and Q.~Q.~Jiang,
P-V criticality and Joule-Thomson expansion of charged AdS black holes in the Rastall gravity,
Chin. Phys. C \textbf{44} (2020) no.6, 065105.

\bibitem{Ade:2015xua}
P.~A.~R.~Ade \textit{et al.} [Planck],
Planck 2015 results. XIII. Cosmological parameters,
Astron. Astrophys. \textbf{594} (2016), A13.


\bibitem{Das:2020boe}
S.~Das, N.~Sarkar, M.~Mondal and F.~Rahaman,
A new model for dark matter fluid sphere,
Mod. Phys. Lett. A \textbf{35} (2020) no.34, 2050280.


\bibitem{DeRisi:2012af}
G.~De Risi, T.~Harko and F.~S.~N.~Lobo,
Solar System constraints on local dark matter density,
JCAP \textbf{07} (2012), 047.


\bibitem{Rahaman:2010xs}
F.~Rahaman, K.~K.~Nandi, A.~Bhadra, M.~Kalam and K.~Chakraborty,
Perfect Fluid Dark Matter,
Phys. Lett. B \textbf{694} (2011), 10-15.


\bibitem{Kiselev:2003ah}
V.~V.~Kiselev,
Quintessential solution of dark matter rotation curves and its simulation by extra dimensions,
arXiv:gr-qc/0303031 [gr-qc].


\bibitem{Akiyama:2019cqa}
K.~Akiyama \textit{et al.} [Event Horizon Telescope],
First M87 Event Horizon Telescope Results. I. The Shadow of the Supermassive Black Hole,
Astrophys. J. \textbf{875} (2019) no.1, L1.

\bibitem{Akiyama:2019eap}
K.~Akiyama \textit{et al.} [Event Horizon Telescope],
First M87 Event Horizon Telescope Results. VI. The Shadow and Mass of the Central Black Hole,
Astrophys. J. Lett. \textbf{875} (2019) no.1, L6.

\bibitem{Kiselev:2004vy}
V.~V.~Kiselev,
Vector field and rotational curves in dark galactic halos,
Class. Quant. Grav. \textbf{22} (2005), 541-558.

\bibitem{Kiselev:2004py}
V.~V.~Kiselev,
Vector field as a quintessence partner,
Class. Quant. Grav. \textbf{21} (2004), 3323-3336.


\bibitem{Li:2012zx}
M.~H.~Li and K.~C.~Yang,
Galactic Dark Matter in the Phantom Field,
Phys. Rev. D \textbf{86} (2012), 123015.

\bibitem{Shaymatov:2020wtj}
S.~Shaymatov, B.~Ahmedov and M.~Jamil,
Reissner-Nordstr\"{o}m-de Sitter black hole immersed in perfect fluid dark matter cannot be overcharged,
arXiv:2006.01390 [gr-qc].

\bibitem{Das:2020yxw}
A.~Das, A.~Saha and S.~Gangopadhyay,
Investigation of the circular geodesics in a rotating charged black hole in presence of perfect fluid dark matter,
arXiv:2009.03644 [gr-qc].

\bibitem{Ma:2020dhv}
T.~C.~Ma, H.~X.~Zhang, H.~R.~Zhang, Y.~Chen and J.~B.~Deng,
Shadow cast by a rotating and nonlinear magnetic-charged black hole in perfect fluid dark matter,
arXiv:2010.00151 [gr-qc].


\bibitem{Narzilloev:2020qtd}
B.~Narzilloev, J.~Rayimbaev, S.~Shaymatov, A.~Abdujabbarov, B.~Ahmedov and C.~Bambi,
Dynamics of test particles around a Bardeen black hole surrounded by perfect fluid dark matter,
Phys. Rev. D \textbf{102} (2020) no.10, 104062.


\bibitem{Hou:2018avu}
X.~Hou, Z.~Xu and J.~Wang,
Rotating Black Hole Shadow in Perfect Fluid Dark Matter,
JCAP \textbf{12} (2018), 040.


\bibitem{Sadeghi:2020xtc}
J.~Sadeghi, E.~N.~Mezerji and S.~N.~Gashti,
Universal relations and weak gravity conjecture of AdS black holes surrounded by perfect fluid dark matter with small correction,
arXiv:2011.14366 [gr-qc].


\bibitem{Zhang:2020mxi}
H.~X.~Zhang, Y.~Chen, P.~Z.~He, Q.~Q.~Fan and J.~B.~Deng,
Bardeen black hole surrounded by perfect fluid dark matter,
arXiv:2007.09408 [gr-qc].

\bibitem{Shaymatov:2020bso}
S.~Shaymatov, D.~Malafarina and B.~Ahmedov,
Effect of perfect fluid dark matter on particle motion around a static black hole immersed in an external magnetic field,
arXiv:2004.06811 [gr-qc].

\bibitem{Hendi:2020zyw}
S.~H.~Hendi, A.~Nemati, K.~Lin and M.~Jamil,
Instability and phase transitions of a rotating black hole in the presence of perfect fluid dark matter,
Eur. Phys. J. C \textbf{80} (2020) no.4, 296.

\bibitem{Rizwan:2018rgs}
M.~Rizwan, M.~Jamil and K.~Jusufi,
Distinguishing a Kerr-like black hole and a naked singularity in perfect fluid dark matter via precession frequencies,
Phys. Rev. D \textbf{99} (2019) no.2, 024050.

\bibitem{Haroon:2018ryd}
S.~Haroon, M.~Jamil, K.~Jusufi, K.~Lin and R.~B.~Mann,
Shadow and Deflection Angle of Rotating Black Holes in Perfect Fluid Dark Matter with a Cosmological Constant,
Phys. Rev. D \textbf{99} (2019) no.4, 044015.

\bibitem{Xu:2017bpz}
Z.~Xu, J.~Wang and X.~Hou,
Kerr\textendash{}anti-de Sitter/de Sitter black hole in perfect fluid dark matter background,
Class. Quant. Grav. \textbf{35} (2018) no.11, 115003.

\bibitem{Xu:2016ylr}
Z.~Xu, X.~Hou, J.~Wang and Y.~Liao,
Perfect fluid dark matter influence on thermodynamics and phase transition for a Reissner-Nordstrom-anti-de Sitter black hole, Adv. High Energy Phys. \textbf{2019} (2019), 2434390.





\end{thebibliography}
\end{document}